\documentclass[conference]{IEEEtran}
\ifCLASSINFOpdf
\else
\fi

\usepackage{cite}
\usepackage{amsmath,amssymb,amsfonts}
\usepackage{algorithmic}
\usepackage{graphicx}
\usepackage{textcomp}
\usepackage{svg}
\usepackage{pdfpages}
\usepackage{hyperref} 
\usepackage{blindtext}
\usepackage{url}
\usepackage{verbatim}
\usepackage[font={small}]{caption, subfig}
\usepackage{adjustbox}
\usepackage{booktabs}
\usepackage{multirow}
\usepackage{colortbl}
\usepackage{indentfirst}
\usepackage{placeins}
\def\BibTeX{{\rm B\kern-.05em{\sc i\kern-.025em b}\kern-.08em
    T\kern-.1667em\lower.7ex\hbox{E}\kern-.125emX}}
\begin{document}
\title{\textit{AGNI}: In-Situ, Iso-Latency Stochastic-to-Binary Number Conversion for In-DRAM Deep Learning}


\author{\IEEEauthorblockN{Supreeth Mysore Shivanandamurthy,
Sairam Sri Vatsavai,
Ishan Thakkar, 
and Sayed Ahmad Salehi}
\IEEEauthorblockA{Department of Electrical and Computer Engineering,
University of Kentucky, Lexington, KY 40506, USA}
% <-this % stops an unwanted space
% \thanks{Manuscript received Jan 19, 2023; revised Jan 19, 2023. 
% Corresponding author: Supreeth M.S. (email: supreethms@uky.edu).}}
}
\IEEEtitleabstractindextext{%
\begin{abstract}
Recent years have seen a rapid increase in research activity in the field of DRAM-based Processing-In-Memory (PIM) accelerators, where the analog computing capability of DRAM is employed by minimally changing the inherent structure of DRAM peripherals to accelerate various data-centric applications. Several DRAM-based PIM accelerators for Convolutional Neural Networks (CNNs) have also been reported. Among these, the accelerators leveraging in-DRAM stochastic arithmetic have shown manifold improvements in processing latency and throughput, due to the ability of stochastic arithmetic to convert multiplications into simple bit-wise logical AND operations. However, the use of in-DRAM stochastic arithmetic for CNN acceleration requires frequent stochastic to binary number conversions. For that, prior works employ full adder-based or serial counter-based in-DRAM circuits. These circuits consume large area and incur long latency. Their in-DRAM implementations also require heavy modifications in DRAM peripherals, which significantly diminishes the benefits of using stochastic arithmetic in these accelerators. To address these shortcomings, this paper presents a new substrate for in-DRAM stochastic-to-binary number conversion called AGNI. AGNI makes minor modifications in DRAM peripherals using pass transistors, capacitors, encoders, and charge pumps, and re-purposes the sense amplifiers as voltage comparators, to enable in-situ binary conversion of input statistic operands of different sizes with iso latency. Our evaluations, based on detailed SPICE simulations (\url{https://github.com/uky-UCAT/AGNI_SPICE.git}), show that AGNI can achieve savings of at least 8$\times$ in area, at least 28$\times$ energy-delay product (EDP), and at least 21$\times$  in $area \times latency$, compared to two in-DRAM stochastic-to-binary conversion circuits from prior works. These circuit-level benefits are demonstrated to propagate at the system-level to achieve at least 3.9$\times$ gain in performance across four deep CNN models. 
\end{abstract}

}

\maketitle

\IEEEdisplaynontitleabstractindextext

% \fi
\IEEEpeerreviewmaketitle

\section{Introduction}
Convolutional Neural Networks (CNNs) have gained immense popularity in recent times and are extensively used in many real-world applications related to machine learning (ML) and Artificial Intelligence (AI) \cite{chen2014dadiannao}\cite{shafiee2016isaac}. These CNNs mimic the structure and behavior of the human brain in visual recognition tasks. In general, CNNs use a large number of computationally complex arithmetic operations such as multiply-accumulate (MAC), nonlinear activation, and pooling. Although these CNN functions can be accelerated due to their high degree of compute parallelism, their acceleration using conventional ASIC platforms (e.g., Dadiannao\cite{chen2014dadiannao}, EIE \cite{han2016eie}) is challenging due to the memory wall problem while accessing their large number of operands \cite{shafiee2016isaac}.

\par In modern deep CNNs, such as RESNET-50\cite{krizhevsky1097imagenet} and GoogLeNet\cite{chris_googlenet}, MAC operations are the most compute and memory intensive operations. Therefore, to accelerate MAC operations, several prior works have explored Processing-In-Memory (PIM) designs. Among these, some PIM designs are based on the emerging non-volatile memory (NVM) crossbar technologies (e.g., ISAAC \cite{shafiee2016isaac}, PRIME \cite{chi2016prime}, XNOR-RRAM \cite{sun2018xnor}), some are based on the traditional DRAM technology (e.g., DRISA \cite{li2017drisa}, SCOPE \cite{li2018scope}, DRACC \cite{deng2018dracc}, LACC \cite{deng2019lacc}), and some are based on the SRAM technology (e.g., \cite{biswas2018conv}\cite{wang201914}\cite{yin2020xnor}). These PIM solutions work to prevent data migration in order to balance memory performance and computational efficiency while processing CNNs locally. 

Among these PIM designs from prior works, the DRAM-based PIM designs are more favorable. This is because, compared to NVM, DRAM is more dominant memory technology for main memory in current computing systems, which makes adopting the DRAM-based PIM accelerators in current computing systems naturally more appealing. Moreover, compared to NVM, DRAM provides lower latency. DRAM is also more tolerant to frequent writing of partial results. On the other hand, SRAM is also prevalent in current computing systems, but the high area cost and low capacity of SRAM makes SRAM-based PIM accelerators less suitable for modern large-scale, deep CNNs. Due to these reasons, DRAM-based PIM accelerators are preferred by the industry as well \cite{lee2021hardware}\cite{ke2021near}. 

%the significant compute requirement and data-intensive operations are MAC(multiply and accumulate) operations. Further, the input computation data on the current cutting edge CNN benchmark applications is massive, such as  with tensor size as large as 15GB. Moreover, using the von-Neumann architecture on this huge amount of data is not energy-efficient and incurs significant latency for the data movement from the CPU and memory\cite{shafiee2016isaac}. 
%\par Several prior works have explored processing-in-memory (PIM) designs based on the emerging non-volatile memory (NVM) crossbar technologies (e.g., ISAAC \cite{shafiee2016isaac}, PRIME \cite{chi2016prime}, XNOR-RRAM \cite{sun2018xnor}) as well as the traditional DRAM technology (e.g., DRISA \cite{li2017drisa}, SCOPE \cite{li2018scope}, DRACC \cite{deng2018dracc}, LACC \cite{deng2019lacc}) to address this data migration length issue. Such PIM solutions work to prevent data migration in order to balance memory performance and computational efficiency while processing CNNs locally.
%\par But, computing MAC operation using the PIM is challenging. Analog NVM crossbar-based PIM CNN MAC accelerators (e.g., ISSAC\cite{shafiee2016isaac} and PRIME\cite{chi2016prime}) require power-inefficient analog-to-digital (ADC) and digital-to-analog (DAC) converters. This additional circuitry indulges in performance degradation and incurs a huge area overhead. Alternatively, digital in-

DRAM-based PIM accelerators for CNNs break a MAC operation into multiple functionally complete memory operation cycles (MOCs). However, these accelerators require huge number of MOCs per MAC, e.g., DRISA requires 222 MOCs per MAC \cite{li2017drisa}. Each MOC can incur up to 49ns latency and consume up to 4nJ of energy. Therefore, to reduce the required number of MOCs per MAC, SCOPE\cite{li2018scope} and ATRIA\cite{shivanandamurthy2021atria} employed stochastic arithmetic. In ATRIA and SCOPE, the use of stochastic arithmetic could reduce the multiplication operations into simple bit-wise logical AND operations, which in turn reduced the per-MAC MOCs to 5/16 for ATRIA \cite{shivanandamurthy2021atria} and 25 for SCOPE \cite{li2018scope}.

Despite these advantages, the use of stochastic arithmetic in ATRIA and SCOPE for CNN acceleration requires frequent stochastic-to-binary (StoB) number conversions; one StoB conversion is required for every point in the per-layer output tensor. In these accelerators, StoB conversions consume substantial latency and energy, even though ATRIA's StoB operations are hidden from the critical path to some extent. SCOPE and ATRIA use a parallel pop counter (Parallel PC) and Serial pop counter (Serial PC)-based StoB conversion, respectively. Parallel PC circuits require full adders, which can take up non-trivial area in DRAM peripherals \cite{kim2015approximate}. On the other hand, Serial PC circuits require high-speed counters. The implementations of full adders and counters in DRAM cannot be optimized for area, latency and energy, due to the constraints of DRAM processes which are significantly different from the standard CMOS logic processes \cite{lenjani2020fulcrum}. As a result, the advantages of using stochastic arithmetic are severely diminished in ATRIA and SCOPE.

To address these shortcomings, this paper presents a new substrate for in-DRAM StoB number conversion called AGNI. AGNI
makes minor modifications in DRAM peripherals using pass
transistors, capacitors, encoders, and charge pumps, to divide the StoB conversion process into four distinct steps: (i) DRAM row activation, (ii) stochastic to analog conversion, (iii) analog to transition-coded unary
conversion, and (iv) transition-coded unary to binary conversion. Each step utilizes a distinct set of DRAM timing signals to orchestrate charge-sharing among various DRAM components, such as sense amplifiers, DRAM cells, bitlines, and capacitors. Moreover, for the "analog to transition-coded unary
conversion" step, AGNI re-purposes the sense amplifiers as voltage comparators. Through all these steps, AGNI enables in-situ binary conversion of input statistic operands of different sizes.

The organization of this paper is as follows: Section II provides preliminaries; Section III describes the structure of our AGNI substrate; Section IV explains the four operational steps of AGNI substrate in detail, using the results of our conducted SPICE simulations (\url{https://github.com/uky-UCAT/AGNI_SPICE.git}); Section V discusses the\textbf{ \textit{overheads}, \textit{SPICE simulations-based and CNN benchmarks-driven performance analysis, error analysis, and comparison with prior works}}, for AGNI; Section VI concludes the paper. 

%Starts with the introduction on StoB conversion in the CNN accelerators and the need for an efficient StoB converter; Next, we provide the background on the drawbacks of a prior StoB converter for in-memory accelerators; Later, architectural overview and hardware modifications are explained; Further, we analyzed the proposed design and provided a comparison result with previous works; Finally, we conclude the paper with future directions.

\section{Preliminaries}
%The widely used StoB converter is the counter-based technique, i.e., serial pop counter \cite{balobas2017high}. This converter works on the principle of cycle-based serial calculation to count the number of 1s in the bit stream. The latency of the operation increases exponentially with the Binary number (BN) length. So, to overcome this drawback, researchers use the parallel pop counter for StoB conversion. Due to the parallel operation of the partial sum creation using the full adders (FAs), latency of the computation is reduced enormously. In parallel PC, the latency has reduced dramatically compared to serial PC. However, the hardware requirement for internal partial sum calculation skyrocketed. Also, these increases in circuitry result in higher energy consumption and area overhead. Further, to overcome the overhead area issue, much research is moving toward the approximate parallel PC. This counter is a hybrid version of parallel PC, but it induces inaccuracy due to the approximation in the calculation. Also, the area overhead is high compared to serial PC \cite{balobas2017high}. There is a need for a novel StoB converter which should be area, energy-efficient with reduced latency for in-memory computing-based deep learning applications. 

\subsection{Stochastic versus Transition-Coded Unary Numbers} \label{transation coding}
\par
\begin{figure}[ht!]
\centering
\includegraphics[scale=0.3]{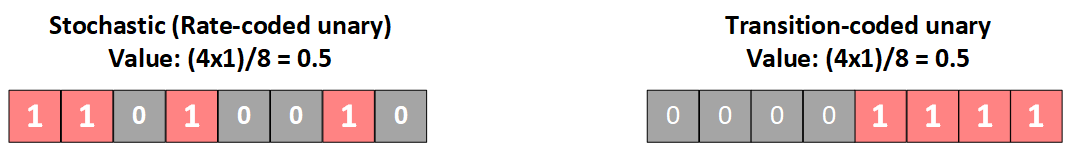}
\caption{The stochastic (rate-coded unary) representation (left) and the transition-coded unary representation (right) of a real value $\upsilon$=0.5.}
\label{fig:1}
\end{figure}

 In the unipolar format of unary computing \cite{wu2020ugemm}, a unary number \textit{W} is a bit-stream of \textit{N} bits that represents a real-valued variable $\upsilon\in[0,1]$ by encoding $\upsilon$  through the ratio $N_1/N$, where $N_1$ is the number of 1's in \textit{W}. As shown in Fig. \ref{fig:1}, a unary number (e.g., $\upsilon$=0.5) can be presented in the stochastic format (also known as rate-coded unary format) (Fig. \ref{fig:1}(left)) or in the transition-coded unary format (Fig. \ref{fig:1}(right)). As evident from the figure, in the stochastic format the '1's in the bit-stream do not appear in a group, whereas in the transition-coded unary format the '1's in the bit-stream appear in group. 

%To clarify the concept of AGNI, we need to give a brief overview of the different unary representations. Due to the brevity, we are restricting the explanation to a unipolar format. Unary computing is broadly classified into rate-coded unary (see Fig.  \ref{fig:1}(a)) and transition-coded unary(see Fig.  \ref{fig:1}(b)). Here in rate coding representation, the information is contained in the frequency of an event. Rate coding is adopted in stochastic computing. The positioning of the 1’s in the bit stream is random. Here in Fig.  \ref{fig:1}(a), the value is 4/8, i.e., 0.5. This probability means seeing one at any bit position is half (0.5). Similarly, in a transition coded scheme where the timing relation of events contains information. Here the 1’s and 0’s are found in groups as shown in Fig.  \ref{fig:1}(b).

\begin{figure}[ht!]
\centering
\includegraphics[scale=0.4]{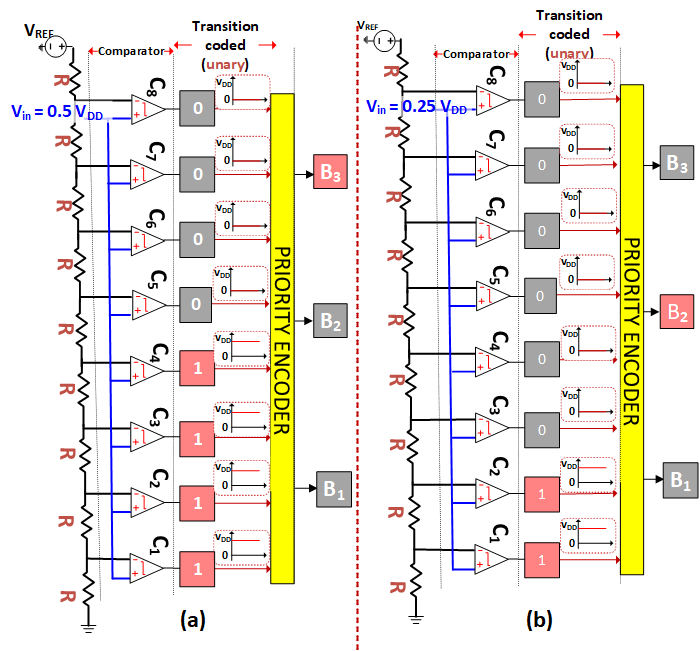}
\caption{Flash ADC with  (a) $V_{in} = 0.5 V_{DD}$, and (b) $V_{in} = 0.25 V_{DD}$.}
\label{fig:2b}
\end{figure}

\subsection{Flash ADC via Transition-Coded Unary Values} \label{flashADC}
% Fig. \ref{fig:2b} shows a schematic of flash ADC with eight inputs. Thus the circuit have eight voltages comparators (i.e.,
% $C_1, C_2, C_3, C_4, C_5, C_6, C_7, and$ $C_8$) as shown in Fig \ref{fig:2b}. The positive terminals of all comparators  connected to the  analog input $V_{in}$. The negative terminal of the comparator is connected to the $V_{REF}$. Now consider the scenario, with inputs equal to 0.5 $V_{dd}$, i.e.
% output detected by the priority encoder is binary digit four as shown in Fig \ref{fig:2b}(a). The highlights in the above ADC is that intermediate data conversion is in rate coding format as shown in Fig. \ref{fig:2b}. 

Fig. \ref{fig:2b} shows a schematic of flash ADC (analog-to-digital converter) with 3-bit binary output. The figure shows the conversion of two example input values. As evident, a B-bit flash ADC employs one voltage divider, a total of 2$^B$ comparators and one 2$^B$:$B$ priority encoder ($B$=3 in Fig. \ref{fig:2b}). Thus, each circuit in Fig. \ref{fig:2b} has eight voltage comparators (i.e., $C_1, C_2, C_3, C_4, C_5, C_6, C_7, and$ $C_8$). The positive terminals of all comparators are connected to the analog
input $V_{in}$. The negative terminal of the comparators are connected to the $V_{REF}$ derived from the resistor-ladder based voltage divider. Suppose a scenario where $V_{in}$ = 0.5 $V_{DD}$ (Fig. \ref{fig:2b}(a)). In this case, the output of the priority encoder is binary four, and upon observation, the output of the comparators that is input to the priority encoder (i.e., the bit sequence 00001111) represents 0.5 in the transition-coded unary format. Similarly, if $V_{in}$ = 0.25 $V_{DD}$ (Fig. \ref{fig:2b}(b)), the output of the priority encoder is binary two, and the output of the comparators (i.e., the bit sequence 00000011) represents 0.25 in the transition-coded unary format. Thus, a flash ADC undertakes analog to binary conversion in two phases: first, analog to unary connversion through the comparators, and second, unary to binary conversion through the priority encoders. \textit{Note that, as discussed later in the paper, our AGNI substrate re-purposes the sense amplifiers in DRAM tiles to implement this first phase of analog to binary conversion.} 

%The above ADC highlights that intermediate data conversion is in rate coding format, as shown in Fig. \ref{fig:2b}.

%Future, if we change the input analog value, the output binary value detected changes. For example, from the initial input ($V_{in}$) value of 0.5  $V_{DD}$ , as in the scenario mentioned above, is changed to 0.25  $V_{DD}$ , then the binary value detected is two as shown in Fig. \ref{fig:2b}(b).

\begin{figure}[ht]
\hspace*{-0.25cm}
\centering
\includegraphics[width=8.5cm,height=8cm]{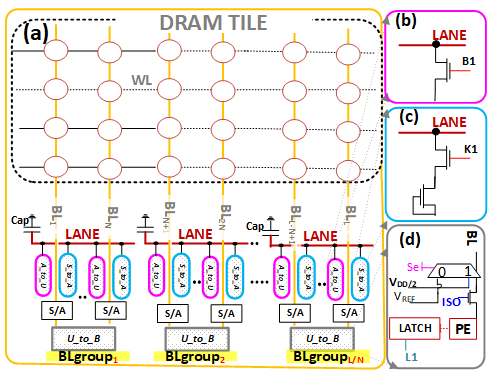}
\caption{Schematic layout of AGNI substrate and employed peripherals. Illustration of (a) an AGNI-modified DRAM tile, (b) an A\_to\_U peripheral unit, (c) an S\_to\_A peripheral unit, and (d) a U\_to\_B peripheral unit.}
%\caption{HERMES architecture overview with tensor mapping for DKV size of 4608 and tensor output count of 512.}
\label{fig:2}
\end{figure}

% \section{Proposed technique to convert the rate coded unary to temporal coded unary}
% This section is useful in better understanding of  AGNI working principle of StoB conversion. Here the input stochastic number is stored and then converted into analog equivalent voltage via ADC as shown in Fig.  \ref{fig:2} using S\_to\_A peripheral unit (will be explained in Section \ref{stoa}). This equivalent analog voltage is converted to transition coded unary by utilizing the A\_to\_U peripheral unit (Section \ref{atou}). Next step is the data extraction of the unary value to compute the number of 1's in bitstream by using priority encoder of U\_to\_B peripheral unit (Section \ref{utob}). 
% \textit{AGNI's novel technique of converting the stochastic to transition coded unary number intermediate step helped by not using the performance in-efficient counter method.}

%begin{figure}[ht]
%\centering
%\includegraphics[scale=0.6]{Fig_1.png}
%\caption{Flow chart for the working principle of AGNI StoB conversion}
%\caption{HERMES architecture overview with tensor mapping for DKV size of 4608 and tensor output count of 512.}
%\label{fig:2}
%\end{figure}

\section{Overview of Our AGNI Substrate} \label{architecutural overview}
% \par The proposed StoB conversion leverage the DRAM circuitry with minimal modification. The functionality of the design is
% discussed in the Section \ref{stages}. However, this section will describe the necessary changes in the AGNI StoB substrate. Majorly, AGNI has four peripheral sections. First, S\_to\_A (stochastic to analog) peripheral unit, each of this unit is dedicated to connected a BL of the DRAM tile as shown in Fig.  \ref{fig:2}(B). Second, A\_to\_U(Analog to unary) peripheral unit, each of this unit is dedicated to connected a BL of the DRAM tile as shown in Fig.  \ref{fig:2} and internal circuitry is shown in Fig. \ref{fig:2}(C). Third, in AGNI substrate the BLs are groups into bitline groups (BLgroup1 to BLgroupN), with each groups consisting of total 'N' BLs. These BLs are connected to a LANE as shown in Fig. \ref{fig:2} with CAP. Fourth, each BLgroup is connected to the U\_to\_B peripheral unit as shown in Fig. \ref{fig:2}. The internal circuitry of the U\_to\_B (unary to binary) unit consist of 2:1 MUXs controlled by SEL (see Fig. \ref{fig:2}(D)). It also comprises of  Priority encounter (PE) and Binary Latch 
% \newline The keypoint in the AGNI substrate is shorter BL of the DRAM subarray. This modification results in a smaller capacitance of \textit{24fF} with a series resistance of $30k\Omega$. 
% Finally, the output temporal coded unary from the A\_to\_U stage of AGNI is converted to binary with a priority encoder and stored into the latch.   

% We call our invented substrate for the in-situ stochastic-to-binary number conversion as AGNI. 
The purpose of our AGNI substrate is to enable in-situ conversion of input stochastic operands (bit-vectors) into binary numbers. To fulfill this purpose, the AGNI substrate employs a few modifications in the structure of each tile of a commodity DRAM module. These modifications are highlighted in Fig. \ref{fig:2}(a). Evidently, our AGNI substrate logically groups the bitlines of each DRAM tile into multiple bitline groups (BLgroups). 
Each BLgroup corresponds to an input stochastic operand. Therefore, the number of bitlines in a BLgroup equals the number of bits in an input stochastic operand (i.e., the size of the input stochastic operand’s bit-vector). Consequently, if the size of each input stochastic operand  is \textit{N} bits, and if each DRAM tile has a total of \textit{L} bitlines (\textit{L} is 256 or 512 typically), then each DRAM tile contains a total of \textit{L/N} logical BLgroups, with each BLgroup having a total of \textit{N} bitlines. 

Further, to enable stochastic-to-binary number conversion of input operands, AGNI employs additional peripherals in each DRAM tile atop the already existing sense amplifiers (SAs). 
As shown in Fig. \ref{fig:2}(a), these peripherals include per-bitline S\_to\_A units, per-bitline A\_to\_U units, per-BLgroup U\_to\_B units, and per-BLgroup analog lanes (see LANEs in the figure). Each LANE is horizontally laid out across a BLgroup and has a capacitor at the end. From Fig. \ref{fig:2}(c), Each S\_to\_A unit contains two transistors (Fig. \ref{fig:2}(c)), whereas each A\_to\_U unit contains one transistor (Fig. \ref{fig:2}(b)). Each U\_to\_B unit contains one isolation transistor (ISO) per bitline (i.e., \textit{N} ISOs per BLgroup) along with one priority encoder (PE) and a latch (Fig. \ref{fig:2}(d)). All S\_to\_A and A\_to\_U units belonging to a BLgroup connect their corresponding bitlines and SAs to the corresponding LANE and analog lane capacitor. The \textit{N} S\_to\_A units of a BLgroup enable stochastic-to-analog number conversion; the \textit{N} A\_to\_U units of the same BLgroup enable analog-to-unary (transition coded unary) number conversion; and the U\_to\_B unit of the same BLgroup enables unary-to-binary number conversion. Thus, the additional peripherals of AGNI enable one stochastic-to-analog-to-unary-to-binary number conversion per BLgroup, thereby enabling a total of \textit{L/N} such conversions in-parallel per DRAM tile. The operation of our AGNI substrate that enables such conversions is discussed next.

\begin{figure}[ht!]
\centering
\includegraphics[width=9cm,height=10.5cm]{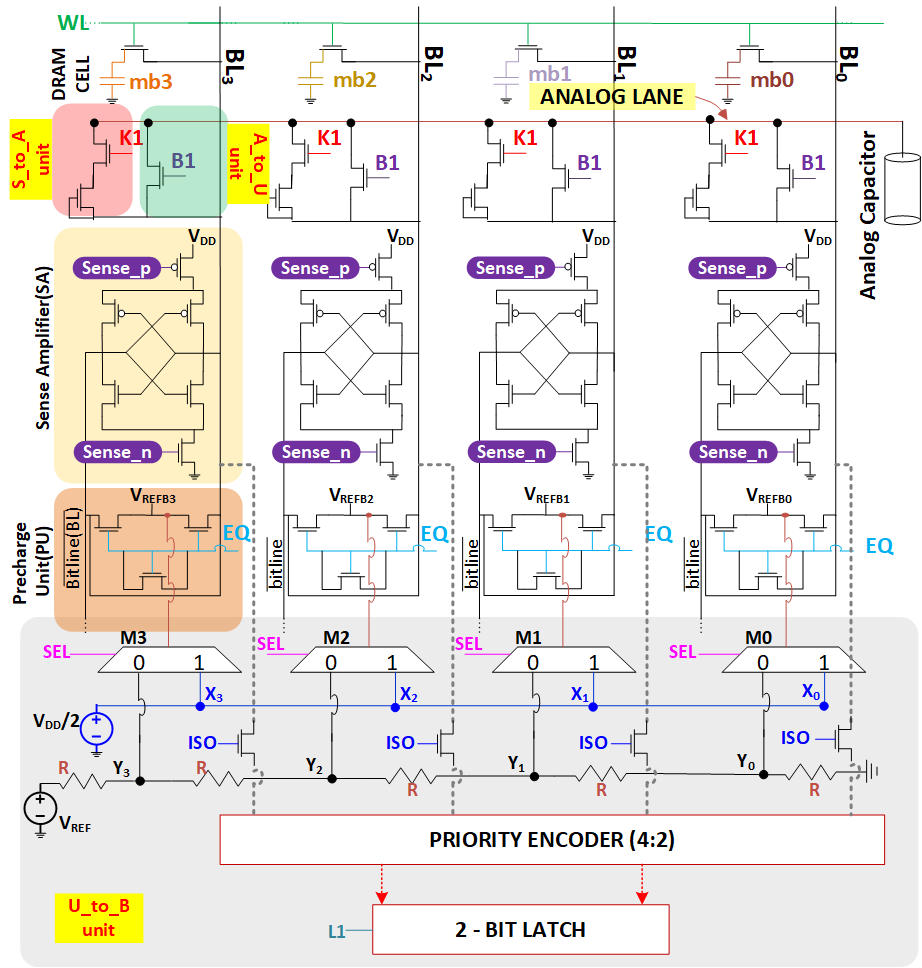}
\caption{Schematic of AGNI substrate for \textit{N} = 4, consisting of peripherals such as S\_to\_A units, A\_to\_U units and U\_to\_B unit.}
%\caption{HERMES architecture overview with tensor mapping for DKV size of 4608 and tensor output count of 512.}
\label{fig:4to2}
\end{figure}

\section{Operation of Our AGNI Substrate}\label{stages}
% As discussed in the above Section \ref{architecutural overview}, StoB operation of AGNI has divided into four stages 1) Reading, 2) S\_to\_A, 3) A\_to\_U, and 4) U\_to\_B. The below subsections will explain each of the mentioned stages in detail. For our design, we are presuming the input rate coded unary value is stored in the DRAM cell (see Fig.  \ref{fig:4to2} mb3, mb2, mb1, and mb0 for SN of length 4 bits). For ease in explanation, consider the operation of AGNI with 4 bits to 2 bits StoB converter with circuitry as shown in Fig.  \ref{fig:4to2}. 

As implied from the previous section, AGNI substrate undertakes stochastic-to-binary conversion of input operands in the following three sequential steps: (i) stochastic to analog (S\_to\_A) conversion, (ii) analog to transition-coded unary  (A\_to\_U) conversion, and (iii) transition-coded unary to binary (U\_to\_B) conversion. For these steps to work for an input stochastic operand, the operand needs to be read into the SAs of its corresponding BLgroup, which can be achieved by activating the DRAM row that contains the stochastic operand. Thus, a DRAM row activation must precede the above three steps, to constitute a sequence of a total of four steps for the operation of AGNI substrate for achieving stochastic-to-binary number conversion. 

To realize these four steps, our AGNI substrate utilizes several timing signals. The timing signals required for the first step (i.e., DRAM row activation) include the standard DRAM operation signals \cite{lee2013tiered} \cite{orosa2021codic}.
The remaining three steps require additional new timing signals to control the added peripherals of the AGNI substrate. The definitions and exact uses of these signals are summarized in Table \ref{AGNI signals}. These signals affect various hardware units of the AGNI substrate. This is illustrated in Fig. \ref{fig:4to2} for an example BLgroup of AGNI substrate with \textit{N} = 4.

\begin{table}[b]
\caption{Definitions and uses of various timing signals employed by AGNI substrate.}
\label{AGNI signals}
\begin{tabular}{|cc|}
\hline
\multicolumn{2}{|c|}{\textbf{Standard DRAM Operation Signals}}                                  \\ \hline
\multicolumn{1}{|c|}{\textbf{WL}} 
& 
\begin{tabular}[c]{@{}c@{}}Signal to turn on a DRAM wordline to\\ enable charge sharing between\\ a row of DRAM cells and corresponding bitlines \end{tabular}                                      \\ \hline

% \begin{tabular}[c]{@{}c@{}}\textbf{sense\_p}\\ \textbf{sense\_n} \end{tabular}
% &
%   \begin{tabular}[c]{@{}c@{}} Complementary signals that are used with each SA to \\enable the sensing and amplification of the\\ bitline voltage perturbation \end{tabular} \\ \hline
\multicolumn{1}{|l|}{\multirow{2}{*}{\begin{tabular}[c|]{@{}c@{}}\textbf{sense\_p}\\ \textbf{sense\_n} \end{tabular}}}
&
  \multicolumn{1}{l|}{\begin{tabular}[c]{@{}c@{}} Complementary signals that are used with each SA to \\enable the sensing and amplification of the\\ bitline voltage perturbation \end{tabular}} \\ \hline

\multicolumn{1}{|c|}{\textbf{EQ}} 
& 
\begin{tabular}[c]{@{}c@{}} Signal for the precharge unit to\\ equalize the BL voltages \end{tabular}\\ \hline
\multicolumn{2}{|c|}{\textbf{Newly Added Timing Signals}}                                       \\ \hline
\multicolumn{1}{|c|}{\textbf{K1}} 
& \begin{tabular}[c]{@{}c@{}}Signal to turn on S\_to\_A units to\\ enable charge flow from the SAs\\ of a BLgroup to the analog LANE capacitor \end{tabular}                    \\ \hline
\multicolumn{1}{|c|}{\textbf{B1}} 
&
  \begin{tabular}[c]{@{}c@{}}Signal to turn on A\_to\_U units to enable charge flow from\\ the analog LANE capacitor of a BLgroup to\\ the bitlines \end{tabular} \\ \hline
\multicolumn{1}{|c|}{\textbf{ISO}} 
&
  \begin{tabular}[c]{@{}c@{}}Signal to turn on/off the isolation transistors, to\\ connect/disconnect the priority encoder from\\ a BLgroup\end{tabular} \\ \hline
\multicolumn{1}{|c|}{\textbf{SEL}} 
&
  \begin{tabular}[c]{@{}c@{}}Signal to the MUXEs that enables the selection of\\ a SA reference voltage ($V_{REF}$)\\ \end{tabular} \\ \hline
\multicolumn{1}{|c|}{\textbf{L1}} 
& Signal to enable the latch for the binary result                            \\ \hline
\end{tabular}
\end{table}

\begin{table}[b]
\centering
\caption{Toggle time stamps ($\uparrow$ or $\downarrow$) for various timing signals to realize the four operational steps of our AGNI substrate.}%  Table showing the signals and time sequences for READ, S\_to\_A, A\_to\_U, and U\_to\_B operation}
\label{timing_signal}
\resizebox{9.25cm}{!}{
\begin{tabular}{lllllllll}
\cline{1-8}
% \multicolumn{1}{|l|}{\multirow{2}{*}{\textbf{\begin{tabular}[c]{@{}l@{}}INITIAL \\ STATE\end{tabular}}}} &
%   \multicolumn{1}{l|}{\textbf{ON}} &
%   \multicolumn{1}{l}{$SEL$} &
%   \multicolumn{5}{c|}{\textit{sense\_p}} &
%       \\ \cline{2-8}
% \multicolumn{1}{|l|}{} &
%   \multicolumn{1}{l|}{\textbf{OFF}} &
%   \multicolumn{1}{l}{K1} &
%   \multicolumn{1}{l}{B1} &
%   \multicolumn{1}{l}{\textit{sense\_n}} &
%   \multicolumn{1}{l}{\textit{ISO}} &
%   \multicolumn{1}{l}{\textit{L1}} &
%   \multicolumn{1}{l|}{\textit{WL}} &
   % \\ \cline{1-8}
\multicolumn{1}{|c|}{\multirow{3}{*}{\textbf{Activate}}} &
  \multicolumn{7}{c|}{\multirow{3}{*}{0ns ${(EQ\uparrow)}$ 5ns ${(EQ\downarrow)}$ 7ns ${(WL \uparrow)}$ 9ns ${(sense\_n\uparrow)}$ 12ns ${(WL\downarrow)}$}} &
   \\
\multicolumn{1}{|c|}{} &
  \multicolumn{7}{c|}{} &
   \\
\multicolumn{1}{|c|}{} &
  \multicolumn{7}{c|}{} &
   \\ \cline{1-8}
\multicolumn{1}{|c|}{\multirow{2}{*}{\textbf{S\_to\_A}}} &
  \multicolumn{7}{c|}{\multirow{2}{*}{13ns ${(K1\uparrow)}$ 37ns ${(K1\downarrow sense\_n\downarrow)}$}} &
   \\
\multicolumn{1}{|c|}{} &
  \multicolumn{7}{c|}{} &
   \\ \cline{1-8}
\multicolumn{1}{|c|}{\multirow{2}{*}{\textbf{A\_to\_U}}} &
  \multicolumn{7}{c|}{\multirow{2}{*}{38ns ${(EQ\uparrow)(SEL\downarrow)}$   42ns ${(EQ\downarrow)}$ 43ns ${(B1\uparrow)}$ 45ns ${(sense\_n\uparrow)}$}} &
   \\
\multicolumn{1}{|c|}{} &
  \multicolumn{7}{c|}{} &
   \\ \cline{1-8}
\multicolumn{1}{|c|}{\multirow{2}{*}{\textbf{U\_to\_B}}} &
  \multicolumn{7}{c|}{\multirow{2}{*}{45ns ${(ISO\uparrow)}$ 51ns ${(L1\uparrow)}$ 52ns ${( L1\downarrow)}$ 55ns ${(B1\downarrow ISO\downarrow)}$}} &
   \\
\multicolumn{1}{|c|}{} &
  \multicolumn{7}{c|}{} &
   \\ \cline{1-8}
 \end{tabular} 
 }
  \end{table}

The BLgroup illustrated in Fig. \ref{fig:4to2} has 4 bitlines, i.e., $BL_0, BL_1, BL_2, and$ $BL_3$. These bitlines correspond to four DRAM bit-cells, i.e., mb0, mb1, mb2, and mb3, respectively. From Fig. \ref{fig:4to2}, each bitline is connected to a SA (highlighted in light yellow) and a pre-charge unit (highlighted in light orange). Additionally, each bitline connects to one S\_to\_A unit (highlighted in light red) and a A\_to\_U unit (highlighted in light green). Moreover, all four bitlines of the BLgroup (i.e., $BL_0, BL_1, BL_2, and$ $BL_3$) connect to one U\_to\_B unit (highlighted in grey), which consists of one \textit{N}:log$_2$\textit{N} priority encoder, one log$_2$\textit{N}-bit latch, \textit{N} isolation transistors (ISOs), one resistor ladder based voltage divider, and \textit{N} multiplexers that select the $V_{REF}$ values for corresponding SAs. A $V_{REF}$ value is either $V_{DD}/2$ or an appropriate level from the voltage divider. The selection of $V_{REF}$ values from the voltage divider enables the SAs to operate as voltage comparators that can provide analog-to-unary conversion (just like flash ADC; Fig. \ref{fig:2b}).
In the following subsections, we explain how the toggling of various timing signals listed in Table \ref{AGNI signals} has to be AGNI signals orchestrated to realize the four operational steps of AGNI substrate.

The exact time stamps for the toggling of these signals are summarized in Table \ref{timing_signal}. The time evolution of these signals are also depicted in Fig. \ref{fig:timing signal}. Note that at the initialization, these signals are in the OFF state. The time evolution of these signals triggers the voltage levels corresponding to various DRAM structures (e.g., bitlines, analog capacitor, bit-cells) to evolve, which is also illustrated in Fig. \ref{fig:timing signal}. We have evaluated various timing and voltage signals depicted in Fig. \ref{fig:timing signal} by modeling and simulating the circuit shown in Fig. \ref{fig:4to2} using LTSpice.

\subsection{DRAM Row Activation (Step 1)}\label{read}
\par
\begin{figure}[ht!]
\centering
\includegraphics[width=9cm,height=14cm]{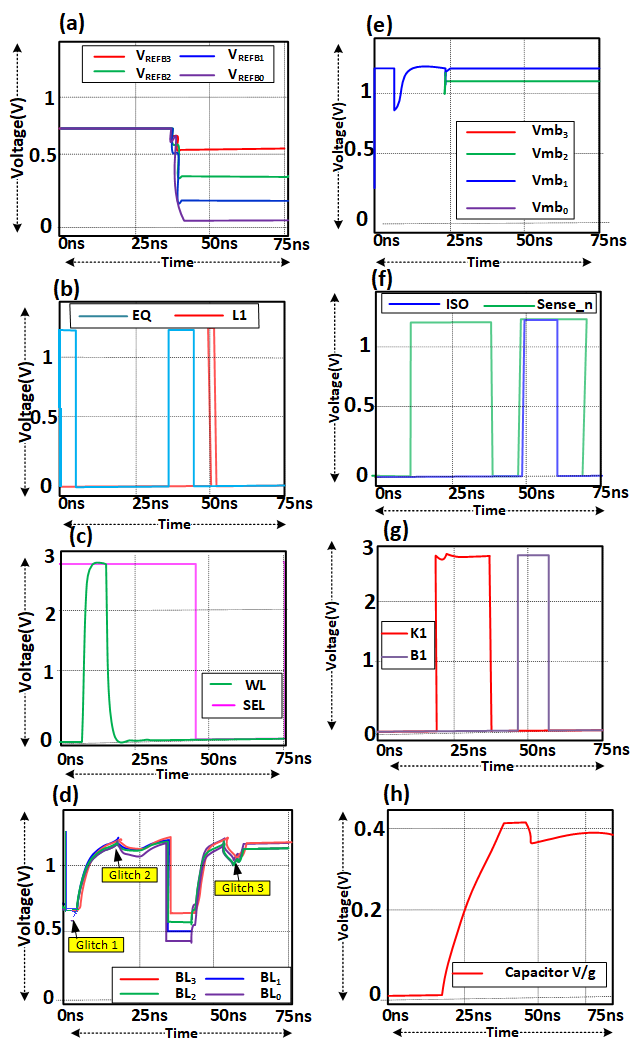}
\caption{Signal evolution traces from SPICE simulations of our AGNI substrate for \textit{N} = 4. (a) voltages of the precharge units ($V_{REF}$), (b) equalizer (EQ) and Latching (L1) signals, (c) wordline (WL) and SEL signals, (d) bitline (BL) voltages, (e) DRAM cell capacitor voltage, (f) sense\_n and isolation (ISO) signals, (g) K1 and B1 signals, and (h) analog capacitor voltage.}
%\caption{HERMES architecture overview with tensor mapping for DKV size of 4608 and tensor output count of 512.}
\label{fig:timing signal}
\end{figure}

 The DRAM row activation step employs EQ, WL, and sense\_n (sense\_p) signals to read the input stochastic operands into the SAs of their corresponding BLgroups. For this step, EQ is first toggled ON (to a higher voltage level in Fig. \ref{fig:timing signal}) at 0 ns. At 0 ns, we consider that SEL has been ON; therefore, at 0 ns, $V_{REF}$ for the SAs has already been selected to be $V_{DD}/2$. As a result, the voltage on the bitlines swiftly settles to $V_{DD}/2$ after 0 ns time-stamp (see the evolution of voltage on the bitlines in Fig. \ref{fig:timing signal}(d)). This step is conventionally known as bitline pre-charging. We are able to achieve swift bitline pre-charging in AGNI because we consider short bitline DRAM architecture with only 8 cells per bitline \cite{lee2013tiered}. After pre-charging, EQ is toggled OFF at 5 ns. Then, at 7 ns, WL is toggled ON. As a result, the DRAM cells (see mb0, mb1, mb2, and mb3 in Fig. \ref{fig:4to2}) connect to their respective bitlines (see $BL_0$, $BL_1$, $BL_2$, and $BL_3$ in Fig. \ref{fig:4to2}) to start sharing their charge with the bitlines. Due to this charge sharing, the voltage on the DRAM cells dips (see Fig. \ref{fig:timing signal}(e) at 7 ns) and the voltage on the bitlines is perturbed (see Fig. \ref{fig:timing signal}(d) at 7 ns). 
%Next, the as mentioned above sequence leads to the incremental voltage change in bitlines ($BL_0, BL_1, BL_2, and$ $BL_3$).

Then, at 9ns, sense\_n (sense\_p) is toggled ON (see Fig. \ref{fig:timing signal}(f)), which enables the SAs to sense the perturbed bitline voltage and amplify it to the full swing. In Fig. \ref{fig:timing signal}(d), since the bitline voltage perturbation is in the positive direction (corresponding to logic '1' stored in the DRAM cells), the bitline voltage is swung to $V_{DD}$ by the SAs. After this full-swing amplification of bitline voltage perturbation, the SAs complete replenishing the DRAM cell voltage at 11 ns. The perturbation amplification and cell replenishing both occur swiftly because we consider the short bitline DRAM architecture for AGNI. Then, at 12 ns, WL is toggled OFF, to disconnect the DRAM cells from the bitlines, to mark the end of the DRAM row activation step. 
%is enabled for sensing . A SA senses this incremental voltage change and pushes to full-swing $V_{DD}$ (if incremental voltage change is +$\delta$V, i.e., DRAM cell is at logic ’1’) or moves to 0 ( if incremental voltage change -$\delta$V, i.e., logic ’0’). 
 
Note, during this step, the bitline voltage evolution encounters transient noise at two events, due to parasitic effects. First, at 5 ns when EQ is toggled OFF. This event is labeled as \textit{glitch 1} in Fig. \ref{fig:timing signal}(d). Second, at 12 ns when WL is toggled OFF (labeled as \textit{glitch 2} in Fig. \ref{fig:timing signal}(d)). 

\subsection{S\_to\_A Conversion (Step 2)}\label{stoa}
% After the DRAM row activation step (\textit{Step 1}), the values are latched into the SA. Now, the
S\_to\_A conversion step (\textit{Step 2}) employs sense\_n (sense\_p), and K1 signals to conduct the conversion of the stochastic operands (which are read into the SAs at the end of \textit{Step 1}) into analog quantities (analog voltage levels). These analog quantities are accrued on respective analog capacitors; one analog voltage level per capacitor per BLgroup. For that, the analog capacitor of each BLgroup is forced to accrue charge incoming from respective SAs by having K1 signal to operate the S\_to\_A peripheral units of the BLgroup. Each S\_to\_A unit consists of a pass transistor and a diode (realized as the back-biased nmos transistor shown in Fig. \ref{fig:2}(c) and Fig. \ref{fig:4to2}). The presence of diode enables the connection of the SA to the pass transistor only if the SA has latched logic '1', i.e., if the corresponding bitline is at $V_{DD}$. Consequently, when K1 signal turns ON the pass transistors of all S\_to\_A units of a BLgroup, the SAs of the BLgroup that are storing $V_{DD}$ connect to the corresponding analog capacitor via the bitlines and analog LANE (Fig. \ref{fig:4to2}). The SAs, if ON, then can force the analog capacitor to accrue some charge, and consequently, some analog voltage level. If SAs are kept ON for a fixed period of time, the accrued analog voltage level on the capacitor would be proportional to the number of SAs that are connected to the capacitor. Since only the SAs that are storing $V_{DD}$ can connect to the analog capacitor, the accrued voltage level on the capacitor would be proportional to the number of logic '1's in the stochastic operand read into the SAs in \textit{Step 1}. This is because only the SAs corresponding to logic '1' bits of the stochastic operand would be storing $V_{DD}$ after \textit{Step 1}). Thus, the voltage level accrued on the analog capacitor provides the analog representation of the stochastic operand. 

Discussing the signal time-stamps, K1 is toggled ON at 13 ns (Fig. \ref{fig:timing signal}(g)). At this time, sense\_n  and sense\_p are already ON, as they were toggled ON during \textit{Step 1} at 9 ns. We then keep K1 and sense\_n (sense\_p) ON for 24 ns, during which the SAs accrue a voltage level on the analog capacitor (see Fig. \ref{fig:timing signal}(h)). In equilibrium, the analog LANE also accrues the same voltage level as the analog capacitor. Then, at 37 ns, both K1 and sense\_n (sense\_p) are toggled OFF (see Figs. \ref{fig:timing signal}(f) and \ref{fig:timing signal}(g)), to mark the end of Step 2. At the end, the voltage level accrued on the analog capacitor and LANE provides the analog representation of the stochastic operand. 

\begin{figure}[b]
\centering
\includegraphics[scale=0.5]{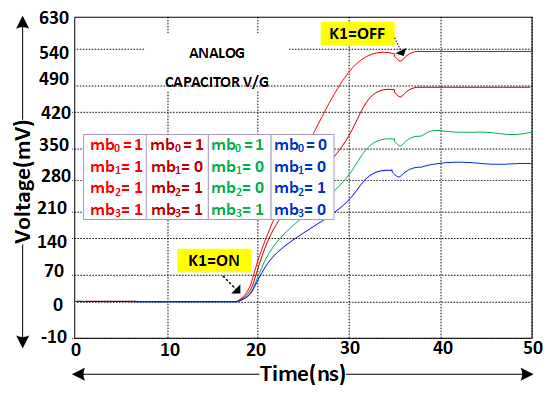}
\caption{Analog capacitor voltage for different 4-bit stochastic numbers.}
%\caption{HERMES architecture overview with tensor mapping for DKV size of 4608 and tensor output count of 512.}
\label{fig: vmax}
\end{figure}

During this step, for how long to keep K1 and sense\_n (sense\_p) ON is really a design choice. But since a thorough exploration of this design choice is beyond the scope of this paper, we decided the duration of 24 ns based on a coarse observation. We observe that our chosen duration of 24 ns is appropriate to provide sufficient noise margin so that different analog voltage levels accrued on the analog capacitor are unerringly distinguishable. We made this observation for our example AGNI substrate with \textit{N}=4 shown in Fig. \ref{fig:4to2}. For \textit{N}=4, the total number of logic '1's in the input stochastic operand can take a total of four distinct values, i.e., 1, 2, 3, and 4. These four distinct values, respectively, correspond to \{mb0=0, mb1=0, mb2=1, mb3=0\}, \{mb0=1, mb1=0, mb2=0, mb3=1\}, \{mb0=1, mb1=0, mb2=1, mb3=1\}, and \{mb0=1, mb1=1, mb2=1, mb3=1\} in Fig. \ref{fig:4to2}. For these, we evaluate how the analog voltage level accrued on the analog capacitor evolves during the 24 ns period; the evolution traces are shown in Fig. \ref{fig: vmax}. For \{mb0=mb1=mb2=mb3=1\}, the accrued voltage at 37 ns reaches the maximum value $V_{MAX}$ = 514 mV. For other cases, it is evident that the accrued voltage level is proportional to the number of '1's in the set \{mb0, mb1, mb2, mb3\} (i.e., in the input stochastic operand). We extend this analysis further and evaluate $V_{MAX}$ for \textit{N} of 16, 32, 64, 128, and 256 (these N values respectively correspond to binary number precision of 4-bit, 5-bit, 6-bit, 7-bit, and 8-bit). The results of $V_{MAX}$ are presented in Table \ref{tab:my-table}. From these results, we observed that that our chosen 24 ns duration was sufficient to provide a total of \textit{N} distinguishable voltage levels on the analog capacitor even for \textit{N}=256. Thus, regardless of the value \textit{N} (i.e., the length of the input stochastic operand), this S\_to\_A step can achieve the analog representation of an input stochastic operand with iso-latency of 24 ns.

\subsection{A\_to\_U Conversion (Step 3)}\label{atou}
As shown in Fig. \ref{fig:2}, a flash ADC undertakes analog to digital (binary) conversion in two stages. In the first stage, an input analog voltage is converted into the equivalent transition-coded unary number using an array of voltage comparators. In the second stage, the unary number is then converted into the corresponding binary number using a priority encoder. The A\_to\_U conversion step of our AGNI substrate implements this second stage of a flash ADC by re-purposing the SAs as comparators. This re-purposing of SAs as comparators is realized using three phases of the A\_to\_U step of our AGNI substrate. For that, signals EQ, SEL, B1, and sense\_n (sense\_p) are employed. 

In the first phase, SEL is toggled OFF at 38 ns, as shown in Fig. \ref{fig:timing signal}(c), to select $V_{REF}$ values from the voltage divider circuit in the precharge units of all \textit{N} SAs of a BLgroup (e.g., see $V_{REFB0}$, $V_{REFB1}$, $V_{REFB2}$, and $V_{REFB3}$ in Fig. \ref{fig:4to2}). Then, in the second phase, EQ is toggled ON at 38 ns and then toggled OFF at 42 ns, to precharge the bitlines of all \textit{N} SAs to their respective $V_{REF}$ values. As a result, between the 38 ns and 42 ns time-stamps, as shown in Fig. \ref{fig:timing signal}(d), the bitline voltages evolve to their respective $V_{REF}$ values. Then, in the next phase, B1 is toggled ON at 43 ns so that the LANE and analog capacitor are connected to the bitlines to enable mutual charge sharing. As a result, the bitline voltages get perturbed by the 45 ns time-stamp (Fig. \ref{fig:timing signal}(d)). If the bitline corresponding to a SA (out of a total of \textit{N} SAs) was precharged to a voltage level greater (less) than the voltage level accrued on the analog capacitor, the perturbation due to charge sharing would increase (decrease) the voltage of that bitline. To sense and amplify this perturbation, sense\_n (sense\_p) is toggled ON at 45 ns. Consequently, the voltages on the bitlines evolve to their full-swing values ($V_{DD}$ or 0V) at 50ns, depending on the direction of the voltage perturbation. Therefore, some of the \textit{N} SAs end up storing logic '1' and the others end up stoting logic '0', and the number of logic '1's out of \textit{N} SAs provides the unary representation of the voltage on the analog capacitor. 

Note that the positions of the '1's in the unary representation differ compared to the positions of the '1's in the input stochastic operand. To understand this, suppose the voltage accrued on the analog capacitor is 0.5$V_{MAX}$ for \textit{N}=4. This would really happen for the case where \{mb0=1, mb1=0, mb2=0, mb3=1\}. In this case, the perturbation would increase the voltages on only $BL_0$ and $BL_1$ in Fig. \ref{fig:4to2}; the voltages on $BL_2$ and $BL_3$ would actually decrease after the perturbation. As a result, the SAs would sense and amplify logic '1's only for $BL_0$ and $BL_1$, thereby providing the positions of '1's in the unary representation to be 0011 (from the left to the right) when the positions of '1's in the stochastic input operand is 1001. This change in the positions of '1's in the unary representation favors the use of the priority encoder to convert from unary to the binary number format. Without this change, converting into the binary number format would require more complex combinational logic, such as a parallel pop counter used in \cite{kim2015approximate}\cite{li2018scope}.

\subsection{U\_to\_B Conversion (Step 4)}\label{utob}
 This step employs ISO, L1 and B1 signals as well as a priority encoder (similar to the one discussed Section \ref{flashADC}), to convert the unary number stored in the SAs at the end of \textit{Step 3} into its corresponding binary number. For that, ISO is toggled ON at 45 ns (Fig.  \ref{fig:timing signal}(f)), so that the ISO transistors turn ON to connect the bitlines to the priority encoder (Fig. \ref{fig:4to2}). Therefore, when the SAs complete evolving the bitline voltages according to their stored unary number at 50 ns (as discussed in \textit{Step 3}), the stored unary number reaches the priority encoder at 50 ns through the bitlines via the ISO transistors. Immediately after that, at around 51 ns, the priority encoder starts providing the converted binary number at its output. Therefore, L1 is toggled ON at 51 ns to enable latching of the priority encoder output (i.e., the binary number result). Then, L1 is toggled OFF at 52 ns, B1 and ISO are toggled OFF at 55 ns, to mark the end of the full operation cycle of AGNI.  
 
% time-stamp, the priority encoder starts so that the  The sequence of this U\_to\_B starts with $ISO$ signal toggle ON (logic 1) at 45ns for a brief of 10ns and ends with toggle OFF at 55ns (see in Fig.  \ref{fig:timing signal}(f)). AGNI wait for 7ns to avoid noise in priority encoder and avoid the inaccuracy. Now the latching sequence signal L1 is initiated for 1ns, i.e, toggle ON at 51ns and ends with toggle OFF at 52ns (see Fig.  \ref{fig:timing signal}(b)). The binary info is latched into the log$_2$\textit{N} bit latch.  At the end of 55ns with toggle OFF of signals (i.e., $B1, ISO$) is enforce. 
 
 Thus, AGNI can convert input stochastic number into the binary format in 55 ns (from \textit{Step 1} to \textit{Step 4}), irrespective of the size of the input stochastic operand (i.e., the value of \textit{N}). Finally, at the end of 55 ns, each BLgroup of AGNI becomes available to convert a new stochastic operand. 
 
 During this step, the bitline voltages experience another glitch at 55 ns time-stamp (labeled as \textit{glitch 3} in Fig. \ref{fig:timing signal}(d)), due to the toggling OFF of B1 that disconnects the bitlines from the LANE and analog capacitor.

\begin{table}[]
\centering
\caption{MAE, MAPE, RMSE, and ${V_{MAX}}$ for AGNI substrate for different BLgroup sizes (different values of \textit{N}). }
\label{tab:my-table}
%\resizebox{\columnwidth}{!}{%
\begin{tabular}{llllllll}
\cline{1-5}

  \multicolumn{1}{|c|}{\textbf{\textit{N}}} &
  \multicolumn{1}{c|}{\textbf{MAE}} &
  \multicolumn{1}{c|}{\textbf{MAPE\%}} &
  \multicolumn{1}{c|}{\textbf{RMSE}} &
  \multicolumn{1}{c|}{\textbf{$V_{MAX}(mV)$}} &
   &
   \\ \cline{1-5}

  \multicolumn{1}{|c|}{16} &
  \multicolumn{1}{c|}{0.28} &
  \multicolumn{1}{c|}{3.58} &
  \multicolumn{1}{c|}{0.41} &
  \multicolumn{1}{c|}{630} &
   &
   \\ \cline{1-5}

  \multicolumn{1}{|c|}{32} &
  \multicolumn{1}{c|}{0.41} &
  \multicolumn{1}{c|}{3.93} &
  \multicolumn{1}{c|}{0.50} &
  \multicolumn{1}{c|}{715} &
   &
   \\ \cline{1-5}

  \multicolumn{1}{|c|}{64} &
  \multicolumn{1}{c|}{0.37} &
  \multicolumn{1}{c|}{1.58} &
  \multicolumn{1}{c|}{1.03} &
  \multicolumn{1}{c|}{735} &
   &
   \\ \cline{1-5}

  \multicolumn{1}{|c|}{128} &
  \multicolumn{1}{c|}{0.29} &
  \multicolumn{1}{c|}{0.97} &
  \multicolumn{1}{c|}{0.43} &
  \multicolumn{1}{c|}{755} &
   &
   \\ \cline{1-5}

  \multicolumn{1}{|c|}{256} &
  \multicolumn{1}{c|}{0.20} &
  \multicolumn{1}{c|}{0.59} &
  \multicolumn{1}{c|}{0.35} &
  \multicolumn{1}{c|}{785} &
   &
   \\ \cline{1-5}
  \end{tabular}%
%}
\end{table}

\section{Evaluation}
\subsection{Overheads of AGNI Substrate}
To evaluate the area overheads of AGNI's peripheral units, we modeled our AGNI substrate on 2D DDR4\_512 DRAM organization at 45 nm technology node using CACTI \cite{ravipati2021fn}. 
Each DRAM cell consumes $6F^2$ area, while the bitline pitch is 3F. Further, the stripes of SAs, precharge units, and write drivers have the heights of \textit{117F, 90F, and 27F} respectively \cite{chang2017understanding} \cite{huang2020improving}. 
Additionally, the heights of the peripheral units of AGNI, such as S\_to\_A, A\_to\_U and U\_to\_B are \textit{27F}, \textit{27F}, and \textit{110F}, respectively. Therefore, the effective height of the AGNI substrate per 2D DDR4 DRAM tile comes out to be \textit{164F}. Therefore, AGNI's total area overhead is $492F^2$. Moreover, we also evaluated the area and power overheads of the charge pump circuits, which we utilized to realize the voltage divider circuit (depicted in Fig. \ref{fig:4to2}) that provides $V_{REF}$ values to the precharge units. We did this evaluation for different values of \textit{N}, using the methods from \cite{jiang2014low}. The results of our evaluation are reported in Table \ref{tab:Charge pump}. 

%In this Section, we also going to explain the area overhead estimate of the peripheral units such as charge pump circuits required for the node voltages to perform the AGNI StoB operations (i.e., A\_to\_U step, S\_to\_U step). For the analysis purpose, we can compute the node voltage generated from the resistance ladder via charge pump (CP) provides the area overhead involved in CP circuits, dynamic power dissipation, area of per charge pump (Acp) referring to \cite{jiang2014low}, and total wasted power per CP for different BLgroups (\textit{N}). From the calculation, for \textit{N} = 256 the total wasted power per CP is $6.85$E-08mW. Similarly, we have provided the values for rest of the BLgroups \textit{N} = 16, 32, 64, and 128.

\begin{table}[]
\centering
\caption{Charge pump (CP) area and power dissipation.}
\label{tab:Charge pump}
\begin{tabular}{|c|c|c|c|}
\hline
\rowcolor[HTML]{FFFFFF} 
\textbf{\textit{N}} &
    \textbf{\begin{tabular}[c]{@{}c@{}}Area of CP\\(Acp)($\mu$$m^2$)\end{tabular}} &
  \textbf{\begin{tabular}[c]{@{}c@{}}Dynamic \\ power per \\CP (W)\end{tabular}} &
  \textbf{\begin{tabular}[c]{@{}c@{}}total \\ wasted power \\ per CP (W)\end{tabular}} \\ \hline
\textbf{16} & 0.0087 & 1.30E-09 & 3.91E-09 \\ \hline
\textbf{32} & 0.0186 & 2.74E-09 & 8.22E-09 \\ \hline
\textbf{64}  & 0.038 & 5.55E-09 & 1.67E-08 \\ \hline
\textbf{128}  & 0.077 & 1.12E-08 & 3.37E-08 \\ \hline
\textbf{256}  & 0.158 & 2.28E-08 & 6.85E-08 \\ \hline
\end{tabular}
\end{table}

\subsection{Setup for Performance Evaluation}
We modeled our AGNI substrate in LTSPICE for $45\eta m$ gpdk technology node for five BLgroup sizes, i.e., with \textit{N} = 16, 32, 64, 128, and 256. We considered the number of bitlines (\textit{L}) per DRAM tile to be 512. \textit{We will make our LTSPICE models publicly available}. For each \textit{N} value, we simulated all possible stochastic numbers as input operands and simulated their conversion into binary numbers using our AGNI substrate's model in LTSPICE. Based on this exercise, we evaluated the mean absolute error (MAE) (Eq. \ref{eq 1}), mean absolute percentage error (MAPE) (Eq. \ref{eq 2}), and Root mean square error (RMSE) (see Eq. \ref{eq 3}) for our simulated stochastic to binary conversions. In these equations (Eqs. \ref{eq 1} to \ref{eq 3}), $y_i$ is the predicted value, $x_i$ is the actual value, and $n$ is the total number of data points. Errors in AGNI substrate mainly emanate from the noise fluctuations during the charge-sharing phases, whenever such fluctuations are larger than the tolerable margins. The resultant error numbers are provided in
Table \ref{tab:my-table}. In addition, as discussed earlier, the table also lists our evaluated $V_{MAX}$  values. 

\begin{equation} \label{eq 1}
    MAE = (\sum_{i=1}^{n}|y_i - x_i|/n)
\end{equation}

\begin{equation} \label{eq 2}
  MAPE = \binom{\underline{1}}{n}\sum_{i=1}^{n}\left| \binom{\underline{x_i-y_i}}{x_i}\right|  
\end{equation}

\begin{equation}\label{eq 3}
    RMSE = \sqrt{\binom{\underline{\sum_{i=1}^{n}(y_i-x_i)^2}}{n}}
\end{equation}

\begin{figure}[h]
\centering
\includegraphics[scale=0.55]{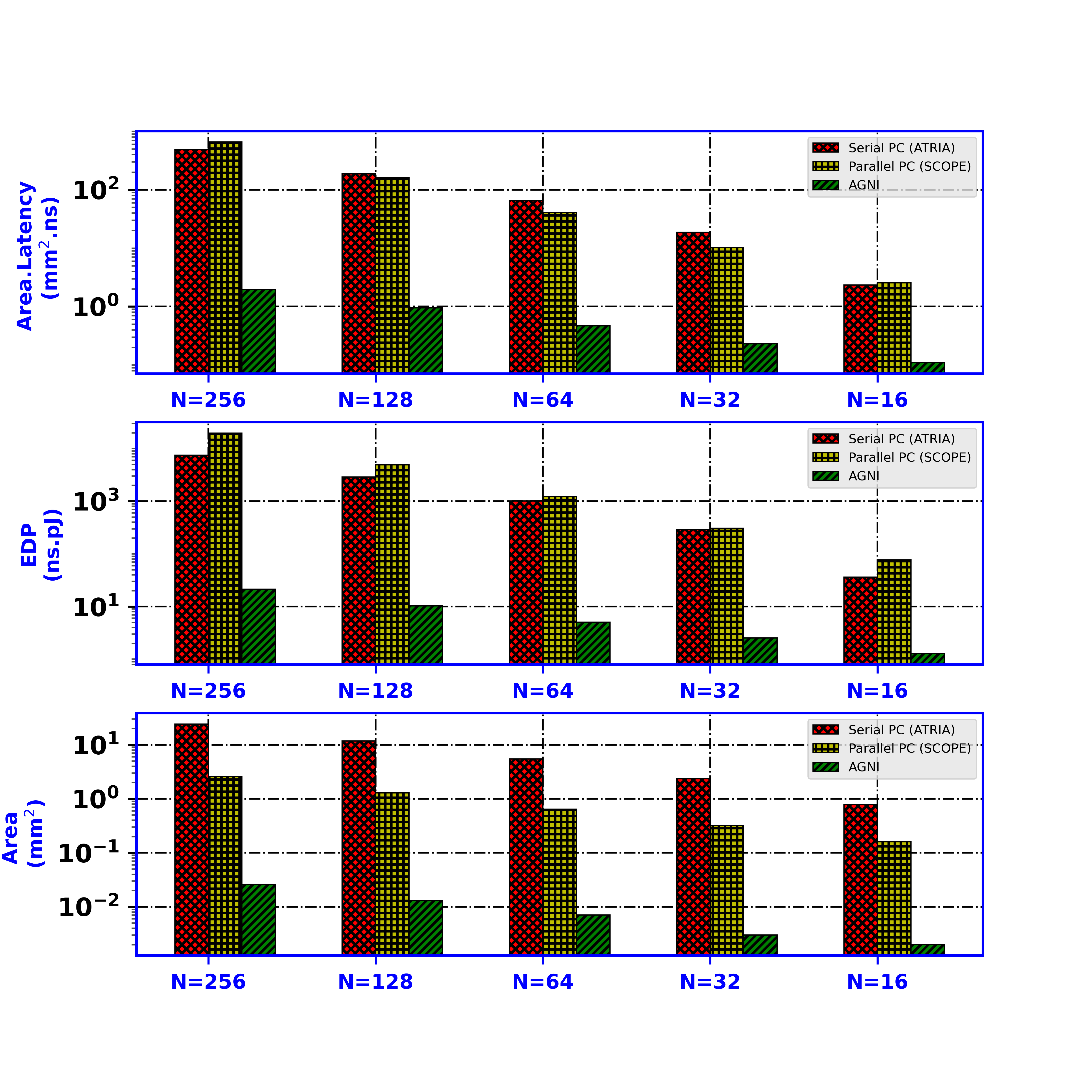}
\caption{$Area \times latency$ (top), energy-delay product (EDP) (middle), and area consumption (bottom) results of prior works (red and yellow columns) and our AGNI substrate (green columns).}
\label{Fig:simulation}
\end{figure}

\begin{figure}[h]
\centering
\includegraphics[scale=0.5]{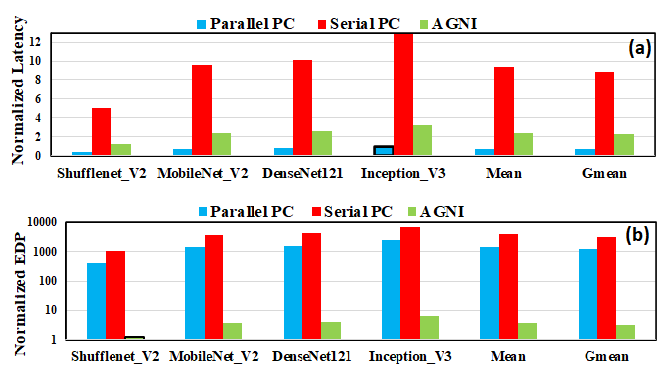}
\caption{System-level results for four CNNs. (a) inference latency normalized to the column for Parallel PC Inception\_V3, (b) inference EDP normalized to the column for AGNI ShuffleNet\_V2. Parallel PC = SCOPE \cite{li2018scope}, Serial PC = ATRIA \cite{shivanandamurthy2021atria}.}
\label{Fig:8}
\end{figure}

We also analyze the performance of our AGNI substrate in terms of area (per BLgroup), energy-delay product (EDP) (per conversion), and $area \times latency$ (per conversion). We compare the results with two stochastic to binary conversion designs from prior work: (1) the parallel pop counter circuit from \cite{kim2015approximate} (referred to as Parallel PC) which is employed by the in-DRAM computing accelerator SCOPE \cite{li2018scope}, (2) the bit-serial pop counter circuit from \cite{kim2015approximate} (referred to as Serial PC) which was employed by the in-DRAM computing accelerator ATRIA \cite{shivanandamurthy2021atria}. The results are given in Fig. \ref{Fig:simulation} (discussed in the next subsection). \textbf{\underline{System-level Evaluation:}} We also leverage our in-house system-level simulator to evaluate the latency (Fig. \ref{Fig:8}(a)) and EDP (Fig. \ref{Fig:8}(b)) for the inference of four CNN benchmarks (i.e., Shufflenet\_V2, MobileNet\_V2, DenseNet121, Inception\_V3) \cite{FChollet2015} for the Imagenet dataset. 
Parallel PC and Serial PC were used to simulate the inference on SCOPE \cite{li2018scope} and ATRIA \cite{shivanandamurthy2021atria} respectively. We only evaluate the StoB phases of these CNNs. 

%To analyze the performance of AGNI StoB in a DRAM converter, we are comparing it with two in-situ popcounters (PCs), i.e., parallel pop counter and serial pop counter. Further, the values tabulated in Table \ref{EDP_table} are considered for in-memory computing accelerators with a parallel pop counter for StoB conversion, used similar to SCOPE \cite{li2018scope}. Likewise, the serial counter’s latency value is considered for in-memory computing such as ATRIA \cite{shivanandamurthy2021atria}. Table III shows the area, Energy delay product (EDP), and $area \times latency$ for the considered prior works and AGNI.

\subsection{Results and Discussion}
The results from Table \ref{tab:my-table} show that AGNI achieves MAE=0.28 for \textit{N}=16 and MAE=0.2 for \textit{N}=256. Similarly, AGNI achieves MAPE=3.58\% for \textit{N}=16 and MAPE=0.8\% for \textit{N}=256. For a given \textit{N}, a total of 2$^N$ different stochastic number values can be represented. Therefore, for a larger \textit{N}, the value $n$ in Eqs. \ref{eq 1} and \ref{eq 2} increases exponentially, which in turn decreases the error magnitudes despite the fact that the decreased tolerance margin at a larger N increases the magnitude of the numerators in Eqs. \ref{eq 1} and \ref{eq 2}.

From Fig. \ref{Fig:simulation}, AGNI achieves $390\times$ less area, $21\times$ less $area \times latency$, and $28\times$ less EDP compared to Parallel PC for \textit{N}=16. For higher values of \textit{N}, AGNI showed significantly greater savings in area, $area \times latency$, and EDP. For example, for \textit{N}=256, AGNI has $923\times$ less area, $247\times$ less $area \times latency$, and $350\times$ less EDP compared to Parallel PC. Parallel PC consumes substantially higher area, $area \times latency$, and energy because it needs to employ full adder circuits \cite{kim2015approximate}, which increases its area and energy consumption. Note that Parallel PC has a slight edge in the latency over AGNI, but this drawback of AGNI can be tolerated for its excellent savings in area, $area \times latency$, and EDP. 

Similarly, from Fig. \ref{Fig:simulation}, AGNI achieves $8\times$ less area, $23\times$ less $area \times latency$, and $59\times$ less EDP compared to Serial PC for \textit{N}=16. For higher values of \textit{N}, AGNI showed significantly greater savings in area, $area \times latency$, and EDP. For example, for \textit{N}=256, AGNI has $96\times$ less area, $333\times$ less $area \times latency$, and $930\times$ less EDP compared to Serial PC. Serial PC performs bit-by-bit counting at a clock rate, which significantly increases its latency and energy consumption compared to AGNI. Moreover, any implementation of a counter logic in-DRAM cannot be optimized for performance or area, due to the constraints of DRAM processes 
\cite{lenjani2020fulcrum}, which increases the area overhead of Serial PC counters compared to the peripherals of our AGNI substrate. Therefore, overall, we observe AGNI to significantly gain in area, $area \times latency$, and EDP, compared to Serial PC. 

\textbf{\underline{System-level Results:}} Fig. 8 shows the normalized inference latency and EDP results for our considered CNNs. 
%and Leveraging our in-house python custom system-level simulator, we calculated the normalized latency (shown in Fig. \ref{Fig:8}(a)) and normalized EDP (shown in Fig. \ref{Fig:8}(b)) of prior work and AGNI on four CNN benchmark applications (i.e., Shufflenet\_V2, MobileNet\_V2, DenseNet121, Inception\_V3). Along with this, we also plotted the arithmetic mean (Mean) and geometric mean (Gmean). 
From the figure, AGNI achieves 3.9$\times$ less latency than Serial PC on Gmean. Further, AGNI achieves 397$\times$ and 1048$\times$ better EDP than Parallel PC and Serial PC, respectively, on average across all considered CNNs. The better EDP results for AGNI confirms its advantages over Parallel PC and Serial PC.  

%Likewise, AGNI achieves  better EDP than Serial PC on average across all considered CNNs. 

% Similarly, we compared our novel design with a serial PC \cite{shivanandamurthy2021atria}. The simulation result shows enormous savings in EDP. Due to the serial counter being cycle based, we observed our designs gained EDP saving of $59\times$ and $930\times$ at 4 bits and 8 bits BN lengths respectively. Further, AGNI also has better area saving over serial PC. We observe $8\times$ and $96\times$ area saving improvement from the simulation at 4 bits and 8 bits BN lengths, respectively.
 
%  Further, from Table \ref{EDP_table}, AGNI has an \textit{$area \times latency$} saving of $21\times$ and $23\times$ w.r.t. Parallel PC and serial PC, respectively, at 4 bits binary operand length. Similarly, at higher BN length, the AGNI’s \textit{$area \times latency$} significantly increases. i.e., $247\times$ and $333\times$ over parallel and serial PC, respectively. Overall, our design has outperformed parallel and serial PC in terms of area and energy.

%For further reference, the simulation files of the AGNI substrate is provided in the following link \href{https://github.com/SupreethMysore/AGNI_SPICE.git}{AGNI simulations link}.

\section{Conclusion and Future Work}

In this paper, we presented a novel DRAM-based substrate called AGNI for in-situ StoB number conversion for Deep learning applications. We discussed the structure and operation of our AGNI substrate in this paper, using the results of our conducted SPICE simulations. We also presented detailed performance analysis results and overheads for our AGNI substrate. Our evaluations show that AGNI can achieve savings of at least 8$\times$ in area, at least 28$\times$ energy-delay product (EDP), and at least 21$\times$  in $area \times latency$, compared to two in-DRAM stochastic-to-binary conversion circuits from prior works. \underline{Future Work}: There is a room for further reducing the latency of AGNI substrate by tightly packing various timing signals of AGNI substrate in a narrower window of time. The capacitance value of the analog capacitor and the physical implementation of the analog LANE also provide avenues for future exploration, to maximize the analog voltage range and noise margin. 

\section*{Acknowledgment}
We thank the anonymous reviewers for their valuable feedback. This research is supported by a grant from NSF (CNS-2139167).

%and metal layer closely The design uses a novel technique of converting the input stochastic bit stream into equivalent analog voltage. This analog voltage is fed into the flash ADC to convert to an unary number. Next, unary value is sensed by the priority encoder to extract the binary value. Converting stochastic to unary mitigate in AGNI, the use of area in-efficient counter-based technique as, in prior works \cite{shivanandamurthy2021atria} is mitigated. The results illustrate that the area overhead for peripheral units such as S\_to\_A, A\_to\_U, and U\_to\_B is minimal i.e., $~4$\% for 8 bits BN length. AGNI has a significant EDP saving of  $350\times$ over parallel PC. These results corroborate the excellent capabilities of AGNI for accelerating the stochastic in-memory accelerator for deep CNNs.

%This paper can be used as a pathfinder to further enhance the StoB conversion, with even better latency and EDP saving. There is lot of scope for improvising the timing and unary conversion method. This research can even be pushed to a deeper sub-nano meter technology node to analyse the performance. 

\bibliographystyle{IEEEtran}
\bibliography{Reference}

\vfill

\end{document}